\documentclass[aps,prl,twocolumn,showpacs,superscriptaddress,groupedaddress,nofootinbib]{revtex4}  
\usepackage{graphicx}  
\usepackage{dcolumn}   
\usepackage{bm}        
\usepackage{amsmath,amssymb}   
\usepackage{aas_macros}
\usepackage{color}
\newcommand{\red}[1]{{#1}}
\def\GeV{{\rm GeV}}

\def\cm{{\rm cm}}

\def\muGs{{\mu \rm Gs}}

\newcommand{\be}{\begin{equation}}
\newcommand{\ee}{\end{equation}}
\newcommand{\ba}{\begin{eqnarray}}
\newcommand{\ea}{\end{eqnarray}}

\begin{document}
\widetext

\title{Testing the axion-conversion hypothesis of 3.5 keV emission with polarization}

\author{Yan Gong}
\affiliation{Key Laboratory of Computational Astrophysics, National Astronomical Observatories, Chinese Academy of Sciences,
20A Datun Road, Beijing 100012, China}

\author{Xuelei Chen}
\affiliation{Key Laboratory of Computational Astrophysics, National Astronomical Observatories, Chinese Academy of Sciences,
20A Datun Road, Beijing 100012, China}
\affiliation{University of Chinese Academy of Sciences, Beijing 100049, China}
\affiliation{Center of High Energy Physics, Peking University, Beijing 100871, China}

\author{Hua Feng}
\affiliation{Department of Engineering Physics and Center for Astrophysics, Tsinghua University, Beijing 100084, China}

\date{\today}

\begin{abstract}

The recently measured 3.5 keV line in a number of galaxy clusters, the Andromeda galaxy (M31) and the Milky Way (MW) center can be well accounted for by a scenario in which dark matter decays to axion-like particles (ALPs) and subsequently convert to 3.5 keV photons in magnetic fields of galaxy clusters or galaxies. We propose to test this hypothesis by performing X-ray polarization measurements. Since ALPs can only couple to photons with polarization orientation parallel to magnetic field, we can confirm or reject this model by measuring the polarization of 3.5 keV line and comparing it to the orientation of magnetic field. We discuss luminosity and polarization measurements for both galaxy cluster and spiral galaxy, and provide a general relation between polarization and galaxy inclination angle. This effect is marginally detectable with X-ray polarimetry detectors currently under development, such as the enhanced X-ray Timing and Polarization (eXTP), the Imaging X-ray Polarimetry Explore (IXPE) and the X-ray Imaging Polarimetry Explorer (XIPE). The sensitivity can be further improved in the future with detectors of larger effective area or better energy resolutions. 
\end{abstract}

\maketitle

\textit{Introduction.} 
A weak emission line at 3.5 keV is detected recently by XMM-Newton and $Chandra$ observations 
in a number of galaxy clusters, as well as in the M31,  and possibly the MW center \cite{Bulbul14,Boyarsky14, Boyarsky15}, though some detections are still debated \cite{Conlon14,Riemer-Sorensen14,Hitomi16}. \red{The origin of this line is still unclear. It might be an unidentified atomic line or lines emitted from the hot ionized intracluster medium \cite{Bulbul14,Gu15,Shan16}.} Another interesting possibility is that it originates from the decay or annihilation of dark matter (such as axion-like particle (ALP), axinos and  sterile neutrinos) to  photons,  (e.g.\cite{Ishida14,Higaki14,Jaeckel14,Abazajian14,Choi14}). However, observations are somewhat puzzling and not entirely consistent with the dark matter direct-production-of-photons interpretation \cite{Bulbul14,Cicoli14}. The 3.5 keV emission from the Perseus cluster is much stronger than other clusters, but the dark matter decay model predicts that the signal should be proportional to dark matter quantity in a cluster. Also, most of 3.5 keV emission comes from the cool cores of clusters, which have much smaller scales than dark matter halos.  These seem to suggest that the line is associated with an unknown astrophysical process in the cool cores of galaxy clusters or galaxies.

Recent works (e.g. \cite{Cicoli14,Conlon14}) proposed that this line is produced by dark matter decay but via an indirect process,  ${\rm DM}\to a\to \gamma$, i.e. dark matter particles first decay to ALPs which have very small mass, and then subsequently the ALPs are converted to photons by interacting with local magnetic field. For example, such decay may happen if the dark matter is a sterile neutrino \cite{Cicoli14,Conlon15}. This model naturally explains why the line is stronger in clusters with cool cores: the conversion probability of ALPs to photons is proportional to $\mathbf{B}_{\perp}^2$, where $\mathbf{B_\perp}$ is the magnetic field perpendicular to the relativistically moving ALP path. In this scenario, the generation of the 3.5 keV X-ray photon is enhanced in regions with strong magnetic field, such as the central region of galaxy clusters with cool cores, or  the central region and spiral arms of spiral galaxies such as M31 and MW \cite{Boyarsky14,Boyarsky15,Cicoli14}.

If the 3.5 keV emission line is generated by this process, how can we confirm it? Here we propose a test experiment by using the X-ray polarization. Only photons with polarization in the direction of magnetic field can couple to ALPs. If the 3.5 keV signal indeed originates from the ALP-photon conversion process,  these photons should be polarized and the orientation of polarization would parallel the magnetic field in the region where they are produced. For observations of nearby galaxies (e.g. M31) and clusters (e.g. Perseus, Coma and Centaurus),  we can identify the 3.5 keV line emission regions, then compare the polarization of the 3.5 keV photons with the magnetic field in that region to check if they are aligned. Other polarization mechanisms are relatively weak  at this wavelength and cannot affect the axion-conversion signal. For distant galaxies and clusters, the individual  emission regions may not be resolvable, but the total strength, orientation and average degree of polarization 
from the entire cluster or galaxy can still be used. 

\textit{Model.} 
In the presence of background magnetic field, the ALP field $a$ is coupled to electromagnetic field via the effective Lagrangian
\be
\mathcal{L}_{a\gamma\gamma} = g_{a\gamma\gamma}~ a ~{\bf E}\cdot{\bf B} \equiv \frac{a}{M}{\bf E}\cdot{\bf B} \, ,
\ee
where $g_{a\gamma\gamma}$ is coupling factor, and the suppression mass scale  $M \gtrsim 10^{10}\GeV$ 
\cite{Olive14}.  
In the presence of a static magnetic field, this interaction term allows the 
conversion of the $a$ particle to a photon with its polarization parallel to $\mathbf{B}$.
For a single domain with homogenous magnetic field, the conversion probability for ALP with energy $\omega$ is given in the linear approximation by \cite{Sikivie83,Raffelt88},
\be \label{eq:Pag}
P_{a\to \gamma} = {\rm sin}^2(2\theta)\,{\rm sin}^2(\frac{\Delta}{{\rm cos}\, 2\theta}).
\ee
Here  tan$\,2\theta=\frac{2B_{\perp}\omega}{M m^2_{\rm eff}}$, and $\Delta=\frac{m^2_{\rm eff}L}{4\omega}$, where $B_{\perp}$ is the magnetic field perpendicular to the direction of motion of ALPs, $L$ is the size of the domain, and $m^2_{\rm eff}=m^2_a-\omega^2_{\rm pl}$ where $\omega_{\rm pl}=\sqrt{4\pi\alpha n_e/m_e}$ 
and $n_e$ is the electron number density. We also have $\Delta/\cos2\theta=\Delta_{\rm osc} L/2$, where
$\Delta_{\rm osc}^2=(\Delta_{\gamma}-\Delta_a)^2+4\Delta_{a\gamma}^2$, with $\Delta_{\gamma}=-\omega_{\rm pl}^2/2\omega$, $\Delta_{a}=-m_a^2/2\omega$, and $\Delta_{a\gamma}=B_{\perp}/2M$. When $2|\Delta_{a\gamma}|\ll |\Delta_{\gamma}-\Delta_a|$, we find $\rm cos\,2\theta \sim 1$, and Eq. (\ref{eq:Pag}) is then reduced to the form used in the study of neutrino oscillation.
Significant production rate is only achieved for the nearly massless ALP case \cite{Cicoli14}, i.e. $m_a \ll \omega_{\rm pl}$.  Thus, for $m_a \approx 0$,  $|\tan2\theta| \sim B_{\perp}\omega/n_eM$ and $\Delta \sim n_eL/\omega$ \cite{Conlon14,Cicoli14}. For typical galaxy or 
galaxy cluster environments,  $n_e \sim 10^{-2}-10^{-3}\ \cm^{-3}$ and $B \sim 10^0-10^1\muGs$, we find 
$|\theta| \ll 1$ and $\Delta\ll 1$ where the small angle approximation applies, and then we have $P_{a\to \gamma} \simeq \frac{1}{4}\left(\frac{B_{\perp}L}{M}\right)^2.$

The magnetic field distribution is found empirically to follow electron number density,  
which takes the form
\be \label{eq:Brz'}
B(r) = B_0 \left[\frac{n_e(r)}{n_0}\right]^p,
\ee
where we assume $p=0.5$ and 1, and $n_0=n_{c0}$ and $n_{g0}$ for galaxy clusters and galaxies, respectively \citep{Cicoli14,Borriello10,Egorov13}. Here $B_0\sim \mathcal{O}(1\,{\muGs})$ for non-cool core galaxy cluster, and $\mathcal{O}(10\,{\muGs})$ for cool core cluster and spiral galaxy \cite{Cicoli14,Beck13,Beck15}. We model electron density with the $\beta$-model for clusters, $n_e(r) = n_{c0} [1+({r^2}/{r^2_c})]^{-3\beta/2}$,
where typical values are  $n_{c0}\sim 10^{-2}\ \rm cm^{-3}$, $r_c\sim 300$ kpc and $\beta\sim 1$.  
The exponential disk 
model \cite{Gomez01} can be used for spiral galaxies, $n_e(r, z') = n_{g0} \exp(-r/r_0)  \exp(-|z'|/z'_0)$, where $r$ is the radial distance 
from the galaxy center in galactic plane, and $z'$ is the vertical height above galactic plane. Typical  values 
are $n_{g0}\sim10^{-2}$ $\rm cm^{-3}$, $r_0\sim15$ kpc, and $z'_0\sim1$ kpc for thick disks, and  $r_0\sim4$ kpc, 
and $z'_0\sim0.05$ kpc for thin disks \cite{Gomez01}. However, we find that the results are quite similar for both cases. 


The photon emission rate from dark matter decay to ALPs followed by ALP-photon conversion is given by $\dot{N}_{\gamma} =N_{\gamma}/\tau_{d\to a}$ where $N_{\gamma}$ is the number of photons, and $\tau_{d\to a}$ is the lifetime of dark matter particles decaying to ALPs, which we assume to be $\tau_{d\to a}\sim \Lambda^2/m_d^3$. For $m_d\simeq7.1$ keV and the dark matter-ALP 
coupling  $\Lambda=10^{17}$ GeV, it is about $10^{27}$s \cite{Cicoli14}.  We assume that the dark matter density distribution follows the Navarro-Frenk-White (NFW) profile \cite{Navarro97}, and the dark matter particle density is denoted as  $n_d(r) = \rho_d(r)/m_d$, where $m_d$ is dark matter particle mass. For galaxy cluster, $N_{\gamma} = \int P_{a\to \gamma}\, n_d(r)\ d^3 r$,  with typical cluster size $R \sim $1 Mpc. For spiral galaxy, we model $N_{\gamma} = D \iint d\phi \, dr\, r \, n_d(r) P_{a\to \gamma}$, where $D$ is the thickness of the layer parallel to the galactic plane.  In the small-angle approximation, the photon production 
rate $\dot{N}_{\gamma} \sim (B_{\perp}L\,m_d/M\Lambda)^2$. 

In the above we have simply integrated the converted photon flux. A rigorous treatment would 
require consideration of the photon-ALP mixing problem, with photoelectric absorption 
by the intervening gas. This can be computed by solving a differential equation for the density matrix, as shown 
in Ref. \cite{Conlon14}. 
For the usual spiral galaxy and galaxy cluster, we find that $P_{\gamma\to a}(r)$ are at most $\sim10^{-3}$ and $\sim 10^{-2}$, respectively.  The photoelectric absorption is at the level of $1\% \sim 10\%$ level depending on the 
inclination angle of the line of sight with respect to the disk. 
Given the current uncertainties on the model parameters, here we neglect such effects.

Next, we estimate the polarization of 3.5 keV photons from galaxy clusters and spiral galaxies. Since cluster magnetic field is irregular \cite{Govoni04}, the total polarization of ALP-converted photons which depend on the magnetic field and properties of member galaxies should vary with different clusters. For typical large clusters with nearly spherical symmetry, we expect the total polarization should be close to zero as the differently polarized photons from different parts of the cluster cancel out. However,  in resolved observations,  we should be able to compare the orientation of magnetic field inside the cluster with the polarization of the photons. If the polarization are aligned with the magnetic field, it would be a strong evidence in support of the ALP-photon conversion scenario,  and vice versa, this scenario can be falsified if it is found otherwise.

\begin{figure}[t]
\begin{center}
\includegraphics[width=0.5\textwidth]{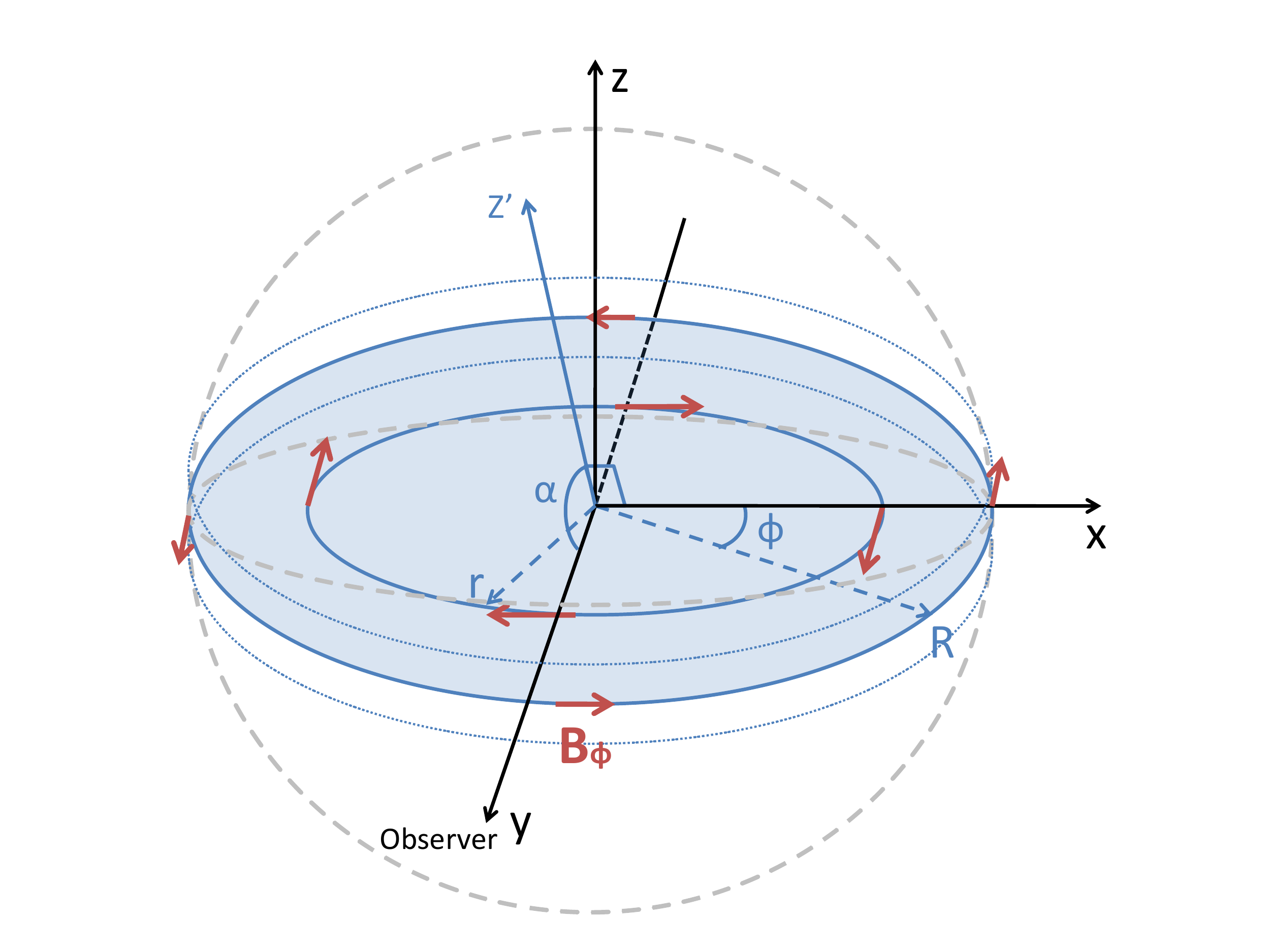}
\includegraphics[width=0.5\textwidth]{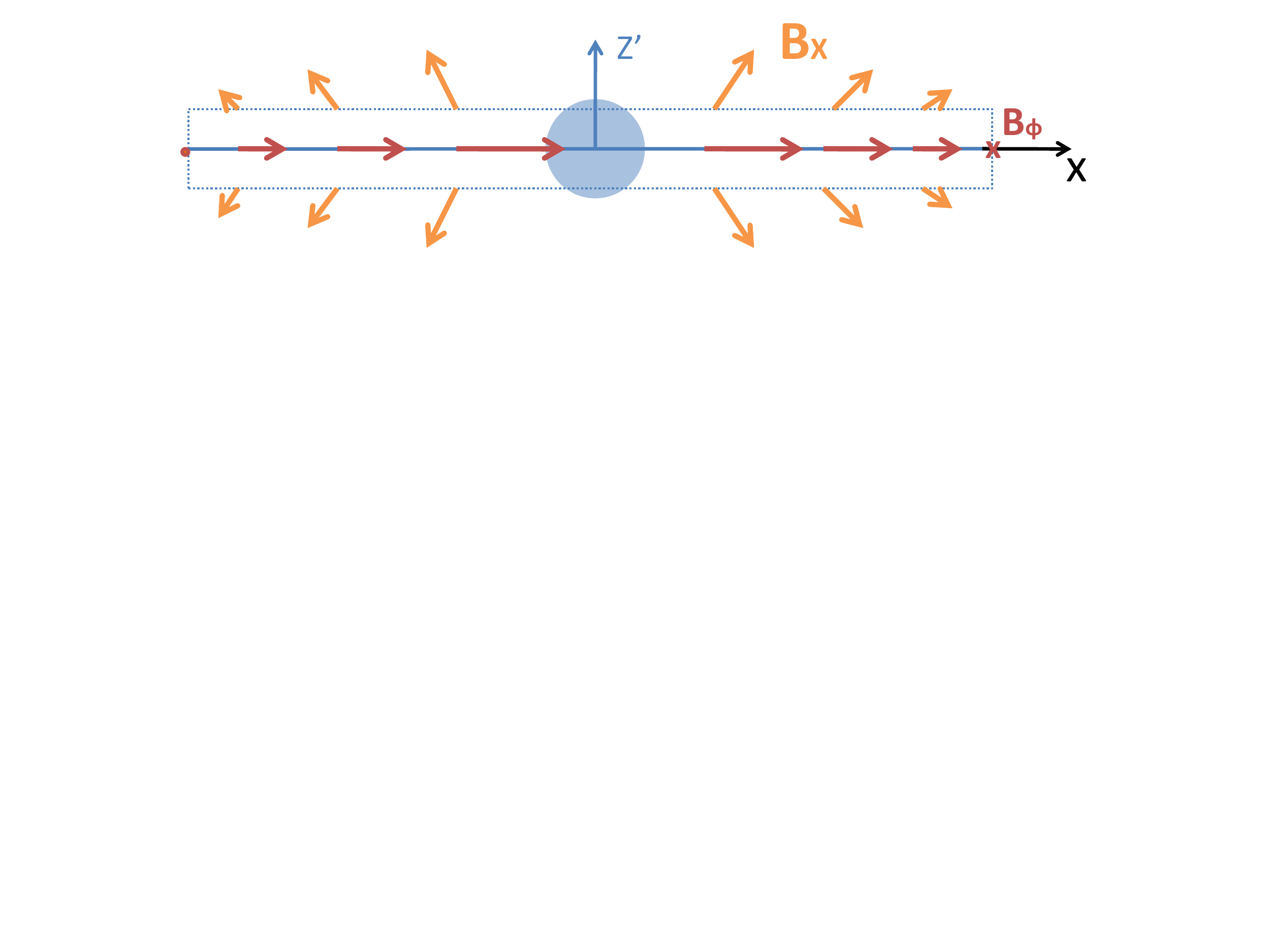}
\end{center}
\vspace{-4.9cm}
\caption{\label{fig:B_model} The magnetic field model for spiral galaxy. The coordinate $y$ is along the line of sight, and $x$ and $z$ axes are along the long and short axis of the projected galactic plane, respectively. The $z'$-axis points to the actual galactic pole, which is inclined with angle $\alpha$, and $\phi$ denotes the polar coordinate on the galactic plane. The dotted lines in both panels denote the middle planes of the two layers above and below the central layer which includes the galactic disk. 
}
\end{figure}

In spiral galaxies, in addition to the small scale stochastic magnetic field, there are large scale regular magnetic fields \cite{Beck13,Beck15}. We consider a two component model of magnetic field: the field within the 
galactic disk which aligns with the spiral arms, and the X-shaped halo field above and below the disk \cite{Beck13,Beck15} (Fig.~\ref{fig:B_model}). The upper panel shows the galactic plane and magnetic field in the disk $B_{\phi}$. For simplicity and as a good approximation in our calculation, we assume $B_{\phi}$ is oriented in circular direction, but have different sense of rotation (clockwise or counter clockwise) between different spiral arms \cite{Beck13,Beck15}. Note that the different sense of rotation of $B_{\phi}$ does not affect our calculation since  the polarization is the same for either case. The lower panel shows the magnetic field $B_X$ above and below galactic disk. Only the $B_X$ close to the disk can affect the ALP-photon conversion, since the field strength $B$ and the coherent length $L$ declines quickly as the galactic latitude increases \cite{Conlon14,Cicoli14}.

We assume the two outer layers has the same thickness $D$ as the central layer which includes the galactic plane, and both $B_{\phi}$ and $B_X$ follow the $n_e$ as Eq.~(\ref{eq:Brz'}), and $B_{\phi}(r) = B(r, z'=0)$,  $B_X = B(r, z'=\pm z'_0)$. We can then compute the projected components $B_{\phi}$ and $B_X$. 
Here we also assume $B_X^{z'}/B_X^r=f_X^{z'r}$, where $B_X^{z'}$ and $B_X^r$ are the $B_X$ components along $z'$ and $r$ directions, respectively. For simplicity $f_X^{z'r}$ is assumed to be a constant at different $r$ \cite{Beck13,Beck15}. The conversion rate can then be estimated with Eq.~(\ref{eq:Pag}) by substituting 
\ba
P_{\phi}^{xz} &=& P_{a\to \gamma}[ B_{\phi}^{xz}, n_e(r,z'=0), L_{\rm eff} ] \\
P_X^{xz} &=& 2\, P_{a\to \gamma}[ B_X^{xz}, n_e(r,z'=z'_0), L_{\rm eff} ].
\ea
The superscript `$xz$' denotes the $x-z$ plane perpendicular to the observer. 
The effective coherent length is given by $L_{\rm eff}=L_d\, {\rm sin}\, \alpha$ where $L_d$ is the coherent length of magnetic field in the galactic plane. If $L_{\rm eff}<D_{\rm disk}$ we set $L_{\rm eff}=D_{\rm disk}$, where $D_{\rm disk}$ is the thickness of galaxy disk. 

The photons of different polarizations are produced at different places so these are incoherent light. The net polarization is given by
\be
f_{\rm Pol} = \frac{N_{\rm gal}^x-N_{\rm gal}^z}{N_{\rm gal}^{xz}},
\ee
 where $N_{\rm gal}^x=N_{\phi}^x+N_X^x$ and $N_{\rm gal}^z=N_{\phi}^z+N_X^z$, and the total number is $N_{\rm gal}^{xz}=N_{\phi}^{xz}+N_X^{xz}$. The degrees of polarization for the central and two outer layers are $ f_{\rm Pol}^{\phi} = (N_{\phi}^x-N_{\phi}^z)/N_{\phi}^{xz}$ and $f_{\rm Pol}^X = (N_X^x-N_X^z)/N_X^{xz}$, respectively. The orientation of polarization is along the $x$ direction (long axis) when $f_{\rm Pol}>0$, and along the $z$ direction (short axis) when $f_{\rm Pol}<0$.

\begin{figure}[t]
\begin{center}
\includegraphics[width=0.47\textwidth]{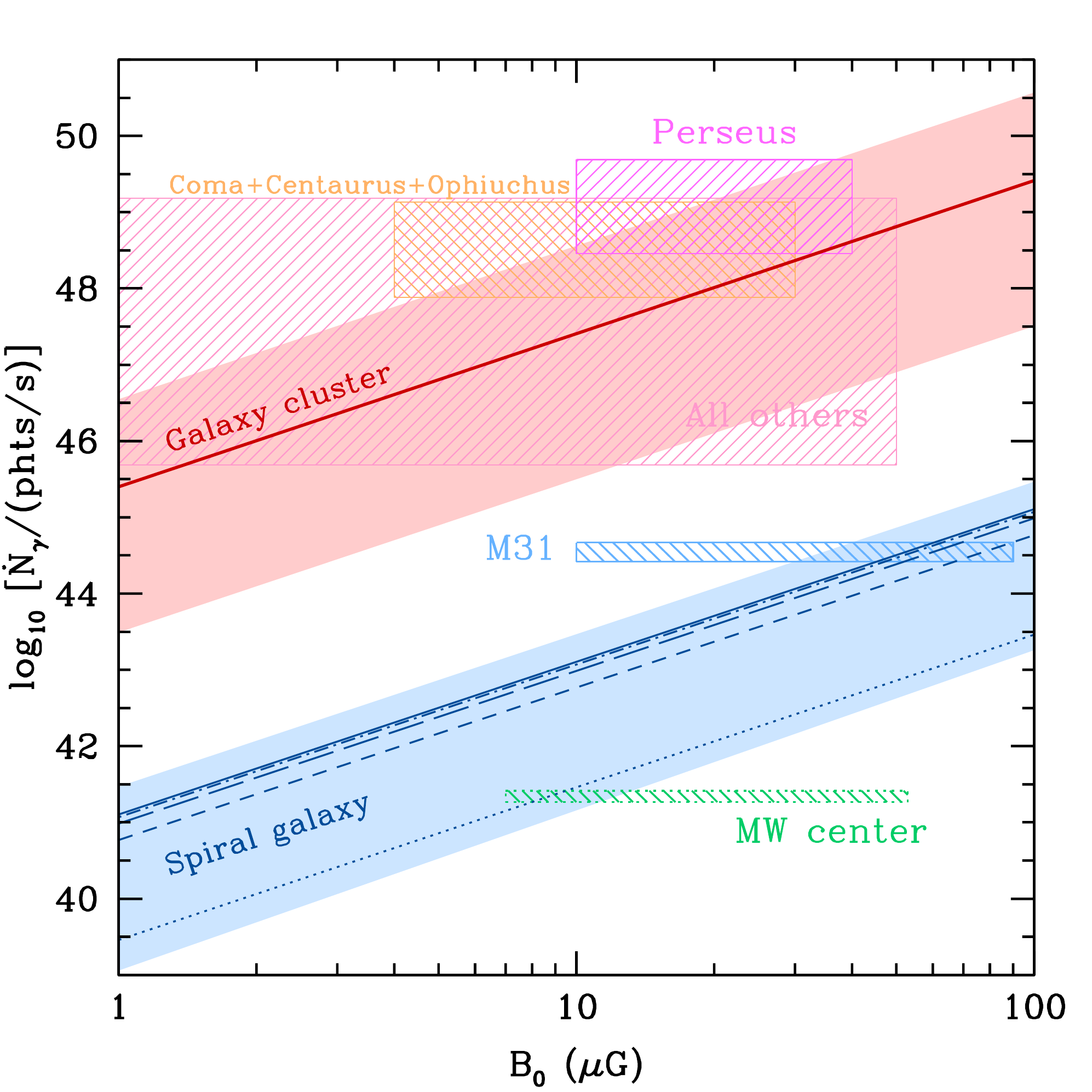}
\end{center}
\vspace{-0.5cm}
\caption{\label{fig:flux} The emission rate  $\dot{N_{\gamma}}$ from ALP-photon conversion as a function of magnetic field $B_0$. The red and blue solid lines show the estimations with magnetic coherent length $L=10$ kpc. The shaded light red and blue regions are derived by taking $L=1 \sim 100$ kpc for clusters and $L=1 \sim30$ kpc for galaxies with inclination angles $\alpha=90^{\circ}$. The blue dash-dotted, long dashed, short dashed and dotted lines are for spiral galaxy with $\alpha= 60^{\circ}$, $ 45^{\circ}$, $30^{\circ}$ and $0^{\circ}$, respectively. We assume the coupling factors $M=10^{13}$ GeV and $\Lambda=10^{16}$ GeV here. The measurements of 3.5 keV emission for clusters and galaxies are also shown as hatched boxes \cite{Bulbul14,Boyarsky14,Boyarsky15}.
}
\end{figure}

\begin{figure}[t]
\begin{center}
\includegraphics[width=0.47\textwidth]{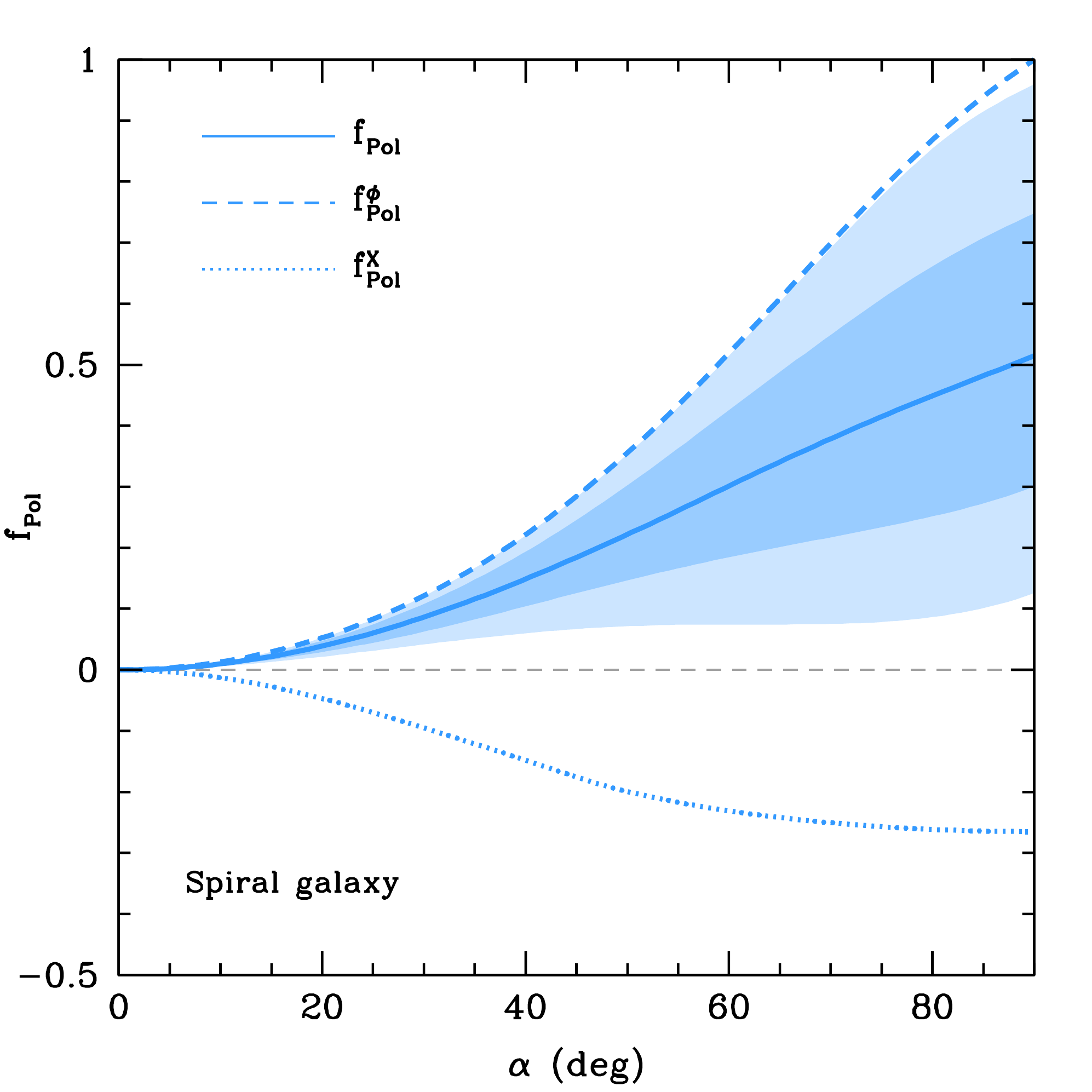}
\end{center}
\vspace{-0.5cm}
\caption{\label{fig:f_pol} Degree of polarization as a function of galactic inclination angle $\alpha$. The $f_{\rm Pol}$, $f_{\rm Pol}^{\phi}$ and $f_{\rm Pol}^X$ are shown in solid, dashed and dotted lines, respectively. The upper and lower bounds of light blue shaded region is obtained with $f_X^{z'r}=0.1$ and $10$, and $f_X^{z'r}=0.5$ and $2$ for deep blue region.
}
\end{figure}

 \textit{Results.} 
 In Figure~\ref{fig:flux}, the predicted emission rate  $\dot{N}_{\gamma}$ is shown as a function of the strength of central magnetic field $B_0$ for galaxy clusters or spiral galaxies. 
On the same plot we also show the measured  3.5 keV emission rates (derived from fluxes \cite{Bulbul14,Boyarsky14,Boyarsky15} and distances) with uncertainties as the hatched regions.  
The flux of cluster 3.5 keV emission are obtained from measurements of the Coma+Centaurus+Ophiuchus clusters and also the stacking result of 69 clusters (marked as ``all others") \cite{Bulbul14}, in the figure we plot the per cluster value obtained by dividing the stacking result by the number of stacked samples. The flux of the M31  \cite{Boyarsky14} and MW \cite{Boyarsky15} are 
$\sim5$ and  $(29\pm5)$ $\times$ $10^{-6}\ \rm phts\ cm^{-2} s^{-1}$ respectively. 
The strongest 3.5 keV line emission is found in the Perseus cluster, with 
$52.0_{-21.3}^{+37.0} \times 10^{-6}$~phts~cm$^{-2}$~s$^{-1}$ \cite{Bulbul14}, though Ref.\cite{Boyarsky14}
found a lower value, and the {\it Hitomi} satellite did not detect it  \cite{Hitomi16}.
These differences in the measured values may be due to the different resolution of the observations. For example, the {\it Hitomi} has  lower angular resolution than the  {\it Chandra} and the {\it XMM-Newton}, so the central AGN and diffuse cluster emissions are mixed. The $Chandra$ data indicates a strong dip in the AGN spectrum around 3.5 keV \cite{Berg16}, so such mixing may give a mild dip at 3.5 keV, which is indeed seen in the {\it Hitomi} data.
Thus, the Perseus 3.5 keV signal could still be real and we assume so below.

 We find that the prediction is generally in good agreements with the observations. The 3.5 keV emission of Perseus is stronger than other clusters as expected, since it is a cool core cluster and has strong central magnetic field as high as $B_0\sim 25\ \mu$G \citep{Taylor06}. The data for the MW center is lower than the prediction, and also several orders of magnitude lower than M31. This may be due partly to the higher magnetic field strength in M31, and partly to the fact that not the whole galaxy but only the central region is measured for the MW. The emission rate from spiral galaxy is much lower than that from galaxy cluster with the same $B_0$, which may explain why there are more detections of 3.5 keV emission in clusters than galaxies \cite{Bulbul14,Boyarsky14, Boyarsky15}.
 
 In Fig.~{\ref{fig:f_pol}}, we plot $f_{\rm Pol}$, $f_{\rm Pol}^{\phi}$ and $f_{\rm Pol}^X$ as a function of galaxy inclination angle $\alpha$ in blue solid, dashed and dotted lines respectively. These results are obtained by taking the ratio $f_X^{z'r}=B_X^{z'}/B_X^r=1$ for the X-shaped magnetic field $B_X$. We also show the results with different $f_X^{z'r}$ in shaded regions. There is no polarization for $\alpha=0^{\circ}$ (face-on galaxy), since the magnetic field in our model is axisymmetric in this case and the polarizations from different parts cancelled out. 
 At $\alpha=90^{\circ} $(edge-on galaxy), $f_{\rm Pol}^{\phi}$ is always equal to $1$, which means  
 photons of ALP-photon conversion from galactic plane are totally 
polarized and the polarization is parallel to the galactic plane. On the other hand, $f_{\rm Pol}^X$ can vary from $-1$ to $1$ at $\alpha=90^{\circ}$ depending on the value of $f_X^{z'r}$. For $f_X^{z'r}=1$, we find $f_{\rm Pol}^X\simeq-0.27$ at $\alpha=90^{\circ}$. The total degree of polarization $f_{\rm Pol}$ is between $f_{\rm Pol}^X$ and $f_{\rm Pol}^{\phi}$. 
\red{Our results are obtained from integration along the line of sight, which is not very 
sensitive to the precise shape of the 
magnetic field. The variations of the result with respect to the inclination angle and strength shown here 
also give some sense of how the results would vary if the shape of the magnetic field varies. }

Note that $f_{\rm Pol}$ is greater than 0 as long as $B_{\phi}$ is larger than $B_X$, which should be the case for most spiral galaxies. Hence, we predict  that the net polarization of the ALP-converted photons is along the long axis of the projected galactic disk for most spiral galaxies. This can be used to test whether the 3.5 keV line originates from this mechanism. 
Of course, if we can measure the inclination angle of a spiral galaxy,  we could also compare the measured $f_{\rm Pol}$ to our prediction. Here we are using a simple magnetic field model,  the predication could be refined if the actually measured value of its magnetic field is available. For example,  
$\alpha=77.5^{\circ}$ for M31, and using the above formula we obtain $f_{\rm Pol}\simeq 0.43$. With the actually measured M31 magnetic field \cite{Fletcher04},  we find $f_{\rm Pol} \sim 0.5$. 

{\it Discussions.}
Sensitive X-ray polarimetry based on the photoelectric effect has become available in recent years~\cite{Costa2001}. Several space missions 
dedicated to or capable of X-ray polarimetry have been proposed and are undergoing phase-A studies, including the eXTP and IXPE/XIPE \cite{Soffitta2013}. 
The energy resolution (FWHM) of the polarization measurement is about $\Delta E=0.8$ keV \cite{Li2015}. 
 Assuming the continuum is unpolarized and the line fully polarized, the observed degree of polarization would be $ f_{\rm pol}^{\rm eff} \approx I_{\rm line}/ S_{\rm cont} \Delta E$, where $S_{\rm cont}$ is the continuum spectral flux density, $I_{\rm line}$ is the line flux, and   $\Delta E$ is the energy resolution.  
 For the Perseus cluster, 
$f_{\rm pol}^{\rm eff}  \sim 0.2\% - 0.6\%$ taken into account of 
 uncertainty. For the 0.6\% polarized case, the signal can be detected with an exposure time of $1.7 \times 10^6$~seconds for eXTP or IXPE/XIPE at a confidence level of 99\% neglecting the instrumental and cosmic X-ray backgrounds.  For the other
 sources, the MW center is most promising case, which needs an exposure time comparable to that of Perseus, while for M31 or the off-center region of Perseus cluster it could be one order of magnitude longer. 

Thus, we see that the polarized signals can possibly be detected with X-ray polarimeters currently under development. 
In the future, X-ray polarimeters equipped on large area telescopes or with better energy resolution or lower systematics will enable us to measure the 3.5 keV line from a large sample of clusters, and test the axion conversion hypothesis.

%
%
%
%
%
%
%
%
%


{\it Acknowledgements.}
We thank Prof. Shuang-Nan Zhang for helpful discussions. YG acknowledges the support of Bairen program from the Chinese Academy of Sciences (CAS). XLC acknowledges the support of the MoST 863 program grant 2012AA121701, the CAS Frontier Science Key Project No. QYZDJ-SSW-SLH017,
the CAS Strategic Priority Research Program  XDB09020301, and the NSFC through grant No. 11373030 and 11633004. HF acknowledges funding support from the NSFC under grant No. 11633003, and the Tsinghua University Initiative Scientific Research Program.


\begin{thebibliography}{61}

\bibitem[]{Bulbul14}
E. Bulbul et al., ApJ, 789, 13 (2014)

\bibitem[]{Boyarsky14}
A. Boyarsky, O. Ruchayskiy, D. Iakubovskyim and J. Franse, PRL., 113, 251301 (2014)

\bibitem[]{Boyarsky15}
A. Boyarsky, J. Franse, D. Iakubovskyim and O. Ruchayskiy, PRL, 115, 161301 (2015)

\bibitem[]{Conlon14}
J. P. Conlon and F. V. Day, JCAP, 11, 033 (2014)

\bibitem[]{Riemer-Sorensen14}
S. Riemer-Sorensen, A\&A 590,  A71

\bibitem[]{Hitomi16}
F. A. Aharonian et al. arxiv:1607.07420

\bibitem[]{Gu15}
L. Gu et al., A\&A, 584, L11 (2015)

\bibitem[]{Shan16}
C. Shah et al., ApJ 833, 52 (2016)

\bibitem[]{Ishida14}
H. Ishida, K. S. Jeong, and F. Takahashi, PLB, 732, 196-200 (2014)

\bibitem[]{Higaki14}
T. Higaki, K. S. Jeong, and F. Takahashi, PLB, 733, 25 (2014)

\bibitem[]{Jaeckel14}
J. Jaeckel, J. Redondo, and A. Ringwald, PRD., 89, 103511 (2014)

\bibitem[]{Abazajian14}
K. N. Abazajian, PRL, 112, 161303 (2014)

\bibitem[]{Choi14}
Choi, K.-Y. and O. Seto, PLB, 735, 92 (2014)

\bibitem[]{Cicoli14}
M. Cicoli, J. Conlon, M. C. D. Marsh, and M. Rummel, PRD, 90, 023540 (2014)

\bibitem[]{Olive14}
K. A. Olive et al. (Particle Data Group), Chin. Phys. C. 38, 090001 (2014).

\bibitem[]{Conlon15}
J. P. Conlon and A. J. Powell, JCAP, 01, 019 (2015)


\bibitem[]{Raffelt88}
G. Raffelt and L. Stodolsky, PRD, 37, 1237 (1988).

\bibitem[]{Sikivie83}
P. Sikivie, PRL, 51, 1415 (1983).


\bibitem[]{Gomez01}
G. C. Gomez, R. A. Benjamin, and D. P. Cox, AJ, 122, 908 (2001)

\bibitem[]{Borriello10}
E. Borriello et al., ApJL, 709, 32 (2010)

\bibitem[]{Egorov13}
A. E. Egorov \& E. Pierpaoli, PRD, 88, 023504 (2013)

\bibitem[]{Beck13}
R. Beck and W. Richard, arxiv:1302.5663.

\bibitem[]{Beck15}
R. Beck, Magnetic Fields in Diffuse Media, Astrophysics and Space Science Library, Volume 407,
Springer-Verlag, 2015, p. 507 (2015), arxiv:1509.04522.

\bibitem[]{Navarro97}
J. F. Navarro, C. S. Frenk, and S. D. M. White, ApJ, 490, 493 (1997).

\bibitem[]{Mirizzi08}
A. Mirizzi, G. G. Raffelt, and P. D. Serpico, Lect. Notes Phys. 741, 115(2008).

\bibitem[]{Govoni04}
F. Govoni and L. Feretti, IJMP, 13, 1549(2004)

\bibitem[]{Taylor06}
G. B. Taylor et al., MNRAS, 368, 1500 (2006)


\bibitem[]{Fletcher04}
A. Fletcher, E. Berkhuijsen, R. Beck, and A. Shukurov, A\&A 414, 53 (2004).

\bibitem[]{Costa2001}
E.~Costa et al., Nature, 411, 662 (2001)

\bibitem[]{Soffitta2013}
P.~Soffitta et al., Exp. Astron., 36, 523 (2013)

\bibitem[]{Li2015}
P.~Soffitta et al., Nuc. Inst. Meth.\ in Phys.\ Res.\ A, 804, 155 (2015)


\bibitem[]{Berg16}
M. Berg et al., arxiv:1605.01043.




\end{thebibliography}
\end{document}